\documentclass[aps,prper,showpacs,twocolumn,floats,superscriptaddress,10pt]{revtex4-1}
\usepackage{graphicx}
\usepackage{dcolumn}
\usepackage{bm}
\usepackage{color}
\usepackage{amssymb}
\usepackage{nicefrac}

\begin{document}
\title{Seebeck effect on a weak link between Fermi and non-Fermi liquids}

\date{\today}

\author{T. K. T. Nguyen}  
\affiliation
{Institute of Physics, Vietnam Academy of Science and Technology,
10 Dao Tan, Hanoi, Vietnam}

\author{M. N. Kiselev}
\affiliation
{The Abdus Salam International Centre for Theoretical Physics, Strada
Costiera 11, I-34151, Trieste, Italy}

\begin{abstract}
We propose a model describing Seebeck effect on a weak link between two quantum systems
with fine-tunable ground states of Fermi and Non-Fermi liquid origin. The experimental realization of the model can be achieved by utilizing the quantum devices operating in the Integer Quantum Hall regime [Z. Iftikhar et al, Nature {\bf 526}, 233 (2015)] designed for 
detection of macroscopic quantum charged states in multi-channel Kondo systems. We 
present a theory of thermo-electric transport through hybrid quantum devices
constructed from quantum dot - quantum point contact building blocks.
We discuss pronounced effects in the temperature and gate voltage  dependence of thermoelectric power associated with a competition between Fermi and Non-Fermi liquid behaviours. High controllability of the device allows to fine-tune the system to
different regimes described by multi-channel and multi-impurity Kondo models.
\end{abstract}

\pacs{73.23.Hk, 73.50.Lw, 72.15.Qm, 73.21.La }

\maketitle

\section{Introduction}

The Fermi-Liquid (FL) theory \cite{Landau} is proven to provide a consistent description of thermodynamic and transport properties of the normal (non-superconducting) metals 
in the presence of weak and strong disorder. In many cases a strong (resonance) scattering of fermions on quantum impurities also mimics the FL properties. 
Majority of quantum impurity models belong to universality classes of a resonant level \cite{Toulous_69} or Anderson/Kondo Hamiltonian \cite{Hewson_book}. 
It is well-known, that Kondo effect \cite{Kondo_1964} at a strong coupling 
limit is described by the local FL paradigm in two important cases: i) fully screened Kondo effect, 
when the number of orbital channels of conduction electrons $M$ is equal to $2S$ ($S$ is a spin of quantum impurity), 
and ii) under-screened Kondo effect when  $M< 2S$. The over-screened limit for $M > 2S$ falls, however, to a completely different universality class known as a Non-Fermi Liquid (NFL)
\cite{Nozieres_Blandin_1980,Affleck,book_bosonization_1998,Hewson_book}.
Significant departure from Fermi Liquid  universality in thermodynamic and transport properties of strongly correlated heavy fermion compounds has been
reported in many experimental works (see. e.g. \cite{Stewart}). Several theoretical models
based on various realizations of the Kondo physics have been utilized \cite{Coleman_book}
for explanation of the strongly-correlated NFL behaviour.

Recently, new experiments in nano-structures \cite{GG_1998, Kouwenhoven_98, Kouwenhoven} lead to a revival of the interest in the Kondo
physics. In addition to shedding a light on understanding the conventional (FL) Kondo physics,
these experiments facilitated an access to the NFL physics \cite{Goldhaber-Gordon_Nature_2007}.  One of the main challenges for engineering the NFL Kondo devices is associated with a lack of stability of the NFL domain under small perturbations created by variation of the external parameters such as magnetic and electric fields. The unstable character of the overscreened Kondo states creates the main obstacle for reliable observation of the NFL physics.
 
In recent experiments \cite{2CK_experiment_nature2015}, a two channel
"charge" Kondo setup has been realized in a single-electron transistor, in which a quantum iso-spin $S=1/2$ is constituted by two degenerate macroscopic charge states of a metallic
island (quantum dot). These experiments not only provide yet another realization of
the "charge" Kondo physics in additional to one theoretically proposed in pioneering works of Flensberg - Matveev - Furusaki (FMF) \cite{Matveev1991,Flensberg,Matveev1995,Furusaki_Matveev}, but also enrich the possibilities of the experimental access to the multi-channel Kondo physics. The "charge" Kondo effect in quantum dot (QD) - quantum point contact (QPC)
semiconductor nano-devices proposed by Matveev and co-workers \cite{Matveev1995,Furusaki_Matveev} relied on counting of mobile electron channels through the quantum number  given by a spin projection 
(and, therefore, the number of channels can only be either one or two as $S=1/2$). The iso-spin-flip processes being a cornerstone of the Kondo physics were related to the 
backward scattering of the quasiparticles (electrons and holes).
The number of channels in the new quantum devices operating in the 
Integer Quantum Hall (IQH) regime \cite{2CK_experiment_nature2015} (c.f. with FMF ideas) is determined by the number of QPC's attached to a metallic QD.
The role of the spin-flip processes in these devices is played by
the electrons' location: either inside the QD (iso-spin up) or outside QD 
(iso-spin down). As a result, firstly, the number of channels in the device can be arbitrary (giving  access to engineering the multi-channel Kondo effect)
and, secondly, the tunability of each QPC provides remarkably high accuracy in experimental  approaching to the unstable NFL strong coupling fixed points.

In our paper we generalize the ideas of FMF theory in applying them to thermoelectric transport. We adopt these ideas to the models describing the new IQH nano-devices and provide a translation of \cite{2CK_experiment_nature2015} onto FMF language. 
We propose a new design for the
quantum dot - quantum point contact (QD-QPC) devices for investigation of weakly coupled Fermi and Non-Fermi liquids (see Fig. \ref{f.1-1}). We develop a theory of thermo-electricity
(Seebeck effect) \cite{Zlatic_book,thermoelectric_coldatomic} for the most intriguing cases of the weak link between two NFLs  and the tunnel contact between FL and NFL.
We demonstrate that the new geometry of quantum devices  not only gives an
access to observation of the NFL fingerprints, but also, thanks to high 
tunability and controllability of the devices, allows to monitor and control
all FL-NFL crossovers.

The paper is organized as follows: We describe the theoretical model for observing the FL and NFL behaviour in Sec. \ref{Sec2}. General equations for the electric conductance 
and thermoelectric coefficients are presented in Sec. \ref{Sec3}. Section \ref{Sec4} is devoted to presentation of the main results. The discussion of the models 
crossover from the weak coupling to the strong coupling, its realization 
with IQH device and possible thermo-transport experiments is given in \ref{Sec5}. The Summary and Conclusions are presented in Sec. \ref{Sec6}.

\section{Model}\label{Sec2}

\begin{figure}[b]
\includegraphics[width=85mm]{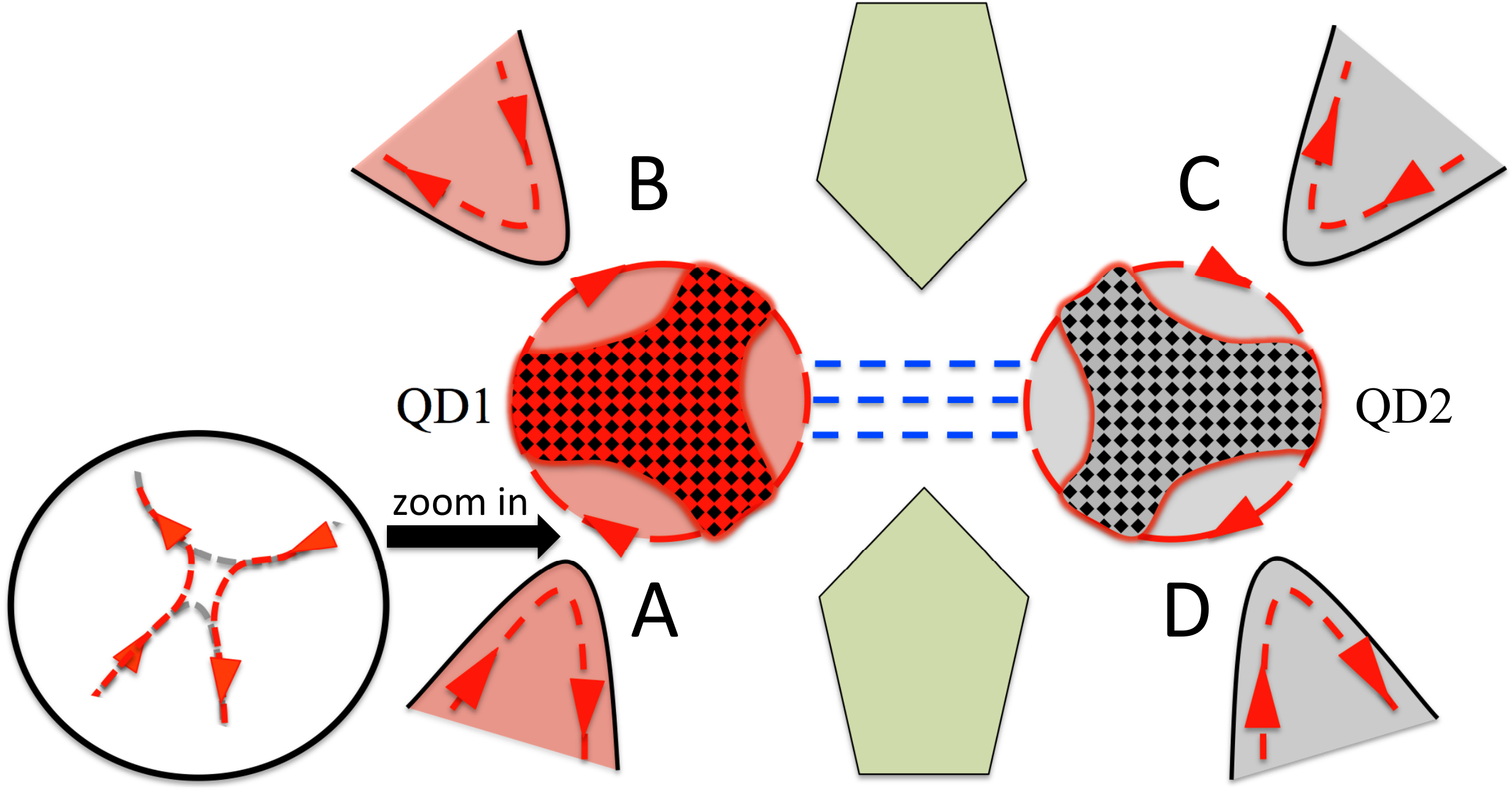} 
\caption{(Color online) Schematic representation of a hybrid metal-semiconductor single electron transistor quantum device:
two quantum dots QD1 and QD2 (cross-hatched areas) are connected through a
tunnel barrier (blue dashed lines) controlled by a split gate (light green boxes). Each QD being a metallic island with continuous electron density of states is electrically coupled
to two-dimensional electron gas (2DEG) denoted by pink and gray areas inside circles and strongly coupled to large electrodes through the quantum point contacts (QPC) labelled by A, B, C, and D. 
The QPCs are fine tuned by field effects in split gates (not shown) to different regimes to provide a weak coupling between i) two coupled Fermi-Liquids;
ii) Fermi-Liquid coupled to a Non-Fermi liquid; iii) two coupled Non-Fermi-Liquids (see Section \ref{Sec4}). The QPCs are formed by the 2DEG (pink and gray areas inside parabolas) placed in perpendicular magnetic field to achieve a regime of the Integer Quantum Hall with $\nu=2$ \cite{2CK_experiment_nature2015}.
The current flows along two spin polarized chiral spin edge channels.
The inner edge channel (not shown) is fully reflected and the outer edge channel 
(shown by red dashed line with arrow) is partially transmitted through 
almost transparent QPCs (see insert showing zoomed in area of QPC). 
The pink color stands for the higher temperature $T+\Delta T$ compared to
the reference temperature $T$ of the gray electrode.}
\label{f.1-1} 
\end{figure}

We consider a setup (see Fig.\ref{f.1-1}) consisting of two large metallic quantum dots
with continuous spectrum  weakly coupled through the tunnel barrier. Each QD is electrically connected (see details in \cite{2CK_experiment_nature2015}) to a two dimensional electron gas (2DEG) denoted by pink and gray areas inside circles and further connected to a large electrode through several quantum point contacts.
The building blocks for a proposed experimental nano-device are the QD-QPC structures used in recent experiments \cite{2CK_experiment_nature2015}. 
The 2DEG confined in the GaAs/Ga(Al)As heterostructure is a subject to strong quantizing magnetic field applied perpendicular to the 2DEG plane. 
The 2DEG is in the Integer Quantum Hall (IQH) regime at the filling factor $\nu=2$. The QPCs are fine-tuned to achieve a regime where the current propagating along the inner chiral edge channel is fully reflected and can be ignored (not shown in Fig.\ref{f.1-1}) while the current propagating along the outer chiral edge channel is partially transmitted across the QPCs (the later is drawn by the dashed red line with the direction shown by the red arrow, gray dashed line on insert Fig. \ref{f.1-1} denotes the current reflected by QPC). 

The logics behind the mapping of IQH setup to a multi-channel Kondo problem is as follows.
Let us consider, for example, the QPC A on Fig.\ref{f.1-1}. The electrons of the electrode A moving along the edge are transmitted at
the QPC A to the pink patch area of the QD1 (which we denote as QD1 A)
with a transmission amplitude close to one. 
Small fraction of electrons is backscattered by the QPC A. We attribute an iso-spin index  (acquiring values $\uparrow$ and $\downarrow$) to the electrons
in QPC A (iso-spin $\downarrow$) and QD1 A (iso-spin $\uparrow$). 
Therefore, the backscattering at QPC 
is equivalent to a spin flip processes. Let's add the QPC B. Using the same logic
we assume that the electrons from the QPC B will be transmitted to the pink patch area 
of QD1 which we denote as QD1 B. Attributing the iso-spin index to the backscattering process at the QPC B we conclude that the number of QPCs is equivalent to the number of orbital channels in the $S=1/2$ Kondo problem. The mapping can be repeated for the QD2. The weak link between QD1 and QD2 (blue dashed lines in Fig.\ref{f.1-1})
connects the pink and gray patch areas (central part of the Figure). The QDs are assumed in the Coulomb blockade regime. Therefore, the total number of electrons in each dot consisting of the number of electrons in color cross-hatched areas and the electrons in all three patches (pink or gray areas confined by the circles) is weakly quantized \cite{Furusaki_Matveev}.

The  Hamiltonian describing two weakly coupled QDs consists of three terms:
$H$$=$$H_{1}+H_{2}+H_{T}$. The Hamiltonian representing each QD (1 or 2) - QPC structure in which the QD is strongly coupled to the leads through QPCs, has the form $H_{j}$$=$$H_{0,j}+H_{C,j}+H_{BS,j}$.

For example, the Hamiltonian $H_{0,1}$ is given by
\begin{eqnarray}
H_{0,1}=iv_{F}\!\!\!\!\sum_{\lambda=\uparrow,\downarrow}\sum_{\alpha=A,B}\!\int_{-\infty}^{\infty}\!\!\!\!\!\!\!\!dx\psi_{\lambda,\alpha,1}^{\dagger}(x)\partial_{x}\psi_{\lambda,\alpha,1}(x).
\end{eqnarray}
Here $\psi_{\lambda, \alpha,1 }$ are the operators describing one-dimensional fermions in the 
QPC$\alpha$ - QD1 system,
$v_F$ is a Fermi velocity, index $\alpha=A,B$ enumerates both QPCs and corresponding patch areas
between QPC$\alpha$ and QD1, index $\lambda$ takes values
$\lambda=\uparrow$ for QD1 A and QD1 B and $\lambda=\downarrow$ for electrodes (A and B). 
We assume that all quantum point contacts are operating in a single mode regime and therefore modelling by 1D fermions is justified \cite{Matveev1991}.

In the same spirit, the Hamiltonian $H_{0,2}$ is written as
\begin{eqnarray}
H_{0,2}=iv_{F}\!\!\!\!\sum_{\lambda=\uparrow,\downarrow}\sum_{\alpha=C,D}\!\int_{-\infty}^{\infty}\!\!\!\!\!\!\!\!dx\psi_{\lambda,\alpha,2}^{\dagger}(x)\partial_{x}\psi_{\lambda,\alpha,2}(x).
\end{eqnarray}

The Hamiltonians $H_{C,j}$ describe the Coulomb interaction in the dots QD1 and QD2 \cite{Ingold_Nazarov}
\begin{eqnarray}
H_{C,j}=E_{C,j}\left[\hat{n}_{\uparrow,j}+\hat n_j-N_{j}({\cal V}_{g,j})\right]^{2}, &  & \:\: j=1,\:2,\label{charging-0}
\end{eqnarray}
where $\hat{n}_{\uparrow,j}=\sum_{\alpha}\psi^\dagger_{\uparrow, \alpha,j}\psi_{\uparrow, \alpha,j}$ denotes the particle number operator of the electrons coming through the QPCs and 
$\hat n_j$ is an integer-valued operator counting the numbers of the electrons coming through the tunnel barrier. Here $E_{C,j}$ are charging energies of the dots and $N_j({\cal V}_{g,j})$ are dimensionless parameters which are proportional to the gate voltages ${\cal V}_{g,j}$.

The Hamiltonian $H_{BS,j}$ models the backward scattering at the
QPCs' positions of the (QD-QPC) $j$  \cite{com0}
{\color{black} (we assume that
the coordinate axes $x_j$ for the left and right devices are chosen independently and therefore left/right QPCs are located at their own origin $x_j=0$).
\begin{eqnarray}
\!\!\!\!\! H_{BS,j} =\!\!\sum_{\alpha}\!\!\int_{-\infty}^{\infty}\!\!\!\!\!\!\!\! dx
\left[\psi_{\uparrow,\alpha,j}^{\dagger}\!\left(x\right)\!V_{\alpha}\!\left(x\right)\!\psi_{\downarrow,\alpha,j}\!\left(x\right)+\text{h.c.}\right].
\end{eqnarray}
}
\noindent
Here $V_{\alpha}(x)$ is a short range QPCs' iso - spin - flip potential.
As we have already pointed out, the QPC index $\alpha=A,B (C,D)$ labels the channels in the multi-channel Kondo problem. Therefore, the model describing a nano-device in the IQH regime (where a
real spin does not play any significant role being fully polarized due to strong magnetic field) is one-to-one mapped to Matveev-Furusaki model
\cite{Matveev1995,Furusaki_Matveev}. 

The tunneling at weak link is described by the Hamiltonian $H_T$:
\begin{eqnarray}
H_{T}=td_{1}^{\dagger}d_{2}+\text{h.c.}\label{h0}
\end{eqnarray}
where $d_{j}$ stands for the electrons in the tunnel patch areas of the dots $j=1,2$
(see Fig. \ref{f.1-1} where pink and gray tunnel patch areas are connected by the blue dashed lines) 
and $t$ is a tunnel amplitude. 

In the spirit of \cite{MAtheory}
we rewrite the $H_{C,j}$ , $H_{0,j}$ and $H_{BS,j}$ in the bosonic language as follows
\begin{eqnarray}
H_{C,j}&=&E_{C,j}\left[\hat{n}_{j}+\frac{1}{\pi}\sum_{\alpha}\phi_{\alpha,j}(0)-N_{j}({\cal V}_{g,j})\right]^{2},\label{charging}\\
H_{0,j}&=&\sum_{\alpha}\frac{v_{F}}{2\pi}\!\!\!\int_{-\infty}^{\infty}\!\!\!\!\!\!dx\left\{\pi^2[\Pi_{\alpha,j}(x)]^{2}+[\partial_{x}\phi_{\alpha,j}(x)]^{2}\right\} ,\label{hr}\\
H_{BS,j}&=&-\frac{\Lambda}{\pi}\sum_{\alpha}|r_{\alpha}|\cos[2\phi_{\alpha,j}(0)], \label{hbs}
\end{eqnarray}
here $\Lambda$ is the bandwidth (ultra-violet energy cut-off), $r_{\alpha}=-iV_{\alpha}(2k_F)/v_F$ is the reflection amplitude of the QPC $\alpha$
on the left side ($j=1$) or the right side ($j=2$), $k_F$ is a Fermi-momentum. 
The field $\phi_{\alpha,j}$
denotes bosonization displacement operator \cite{book_bosonization_1998}  describing transport through
the QPC $\alpha, j$ with a scatterer at $x_j=0$, and $\Pi_{\alpha,j}$
is the conjugated momentum $[\phi_{\alpha,j}(x),\Pi_{\alpha',j}(x')]=i\pi\,\delta(x-x')\delta_{\alpha\alpha'}$ \cite{nkk_2010}.

{\color{black} As it was shown in detail in \cite{Matveev1991,Furusaki_Matveev,Sela,LeHur,LeHur_Seelig}, the FMF theory modelling the QD-QPC structure with spin can be 
mapped to the two channel Kondo model as }
{\color{black}
\begin{eqnarray}
H_{K,j}=\sum_{\alpha}
J^{\alpha,j}_{\perp}\left(\psi_{\uparrow,\alpha,j}^{\dagger}(0)
\psi_{\downarrow,\alpha,j}(0)\hat{S}^{-}_{j}+\text{h.c.}\right)+\Delta E_j \hat{S}^z_j,\nonumber
\end{eqnarray}
where  $\hat{S}^{\pm}_j=\hat{S}^x_j\pm i\hat{S}^y_j$, ${S}^\pm_j$ accounts for adding and subtracting one electron to the dot $j$. The Kondo coupling parameter $J^{\alpha,j}_{\perp}$ is proportional to the reflection amplitude $r_{\alpha}\!\left(x=0\right)$ of the QPC $\alpha$ and
$\Delta E_j=E_{C,j}(1-2N_j)$ is the energy splitting of the $n_j=0$ and $n_j=1$ states of the dot.
}

{\color{black}
The transverse part of the Kondo Hamiltonians $H_{K,1}$ and $H_{K,2}$ are
straightforwardly  bosonized \cite{LeHur_Seelig}. For simplicity we present the bosonized form of $H_{K,1}$. Similarly, $H_{K,2}$ is obtained by replacing indexes $A\to C$ and $B\to D$ and $1\to 2$:
\begin{eqnarray}
H_{K,1}\!&=&\!\!
\frac{\Lambda}{\pi}
\left[\!J_{x}^{(1)}\!\cos[\!\sqrt{2}\phi_{s,1}(0)]\hat{S}^x_{1}\!+\!\!J_{y}^{(1)}\!\sin[\!\sqrt{2}\phi_{s,1}(0)]\hat{S}^y_{1}\!\right]\nonumber, \label{hbsK}
\end{eqnarray}  
in which we i) keep only the iso-spin mode $\phi_{s,1}(x)=[\phi_{A,1}(x)-\phi_{B,1}(x)]/\sqrt{2}$ and the same time neglect the charge mode $\phi_{c,1}(x)=[\phi_{A,1}(x)+\phi_{B,1}(x)]/\sqrt{2}$; 
ii) define the Kondo coupling parameters as follows:
$J_{x}^{(1)}= 2v_F\sqrt{\gamma E_{C,1}/\Lambda}||r_{A}|+|r_{B}||\cos(\pi N_1)$ and 
$J_{y}^{(1)}= 2v_F\sqrt{\gamma E_{C,1}/\Lambda}||r_{A}|-|r_{B}||\sin(\pi N_1)$ ($J^{(1)}_{x}\propto J^{A,1}_{\perp}+J^{B,1}_{\perp}$, $J_{y}^{(1)}\propto J^{A,1}_{\perp}-J^{B,1}_{\perp}$), $\gamma=e^{\textbf{c}}\approx 1.781$, $\textbf{c}\approx 0.577$ is Euler's constant.}

\section{Thermoelectric transport through a weak link}\label{Sec3}

In order to study the Seebeck effect in the device shown on Fig. \ref{f.1-1}, the right QD2-QPCs part (the drain) is prepared at the reference temperature $T$. 
The left QD1-QPC (the source) is heated by the current heat technique \cite{current_heat}
to achieve a small temperature drop $\Delta T$ across the tunnel barrier separating QD1 and QD2. The temperature drop $\Delta T$ is assumed to be small compared to the reference temperature $T$  to guarantee the linear response operation regime for the device \cite{MAtheory}. 

We sketch the derivation of the charge current and the thermoelectric coefficients 
assuming a weak link (tunnel barrier) separating two QD-QPC nano-devices 
(see Fig \ref{f.1-1}).
The current $I_{\rm sd}$ in lowest order in tunneling amplitude $t$ is given by: 
\begin{equation}\label{tuncur}
I_{\rm sd}=-2\pi e|t|^{2}\!\!\!\int_{-\infty}^{\infty}\!\!\!\!\!\!d\epsilon\:\nu_{1}(\epsilon)\nu_{2}(\epsilon)\left[f_{1}(\epsilon)-f_{2}(\epsilon)\right]~.
\end{equation}
The tunnel density of states (DoS) in the left [$\nu_1(\epsilon)$]  and right [$\nu_2(\epsilon)$] QD-QPC devices are expressed in terms
of the Matsubara Green's Function as follows:
\begin{equation}
\nu_{j}(\epsilon) = -\frac{1}{\pi}\cosh\left(\frac{\epsilon}{2T}\right)\int_{-\infty}^{\infty}\!\!\! d\tau {\cal G}_{j}\left({\displaystyle \frac{1}{2T}+i\tau}\right)e^{i\epsilon \tau}~.
\end{equation}
Here ${\cal G}_{j}(\tau)$ ($j=1, 2$) are exact Green's Functions (GF) of interacting fermions and $f_{1}(\epsilon)=f(\epsilon+e\Delta {\cal V},T+\Delta T)$,
 $f_{2}(\epsilon)=f(\epsilon,T)$ are corresponding distribution functions. 

Following Matveev-Furusaki \cite{Furusaki_Matveev} we introduce a counting operator $F_j$:
$\psi_{\uparrow,\alpha,j}(-\infty) \to  
\psi_{\alpha,j}^{(0)}\hat{F}_{j}$ 
in order to account for effects of Coulomb interaction in each QD and reflection at QPCs.
The operators $F_j$ obey the commutation relation $[$$\hat{F}_{j}$$,$$\hat{n}_{j}$$]$$=$$\hat{F}_{j}$ \cite{Ingold_Nazarov}. We define the GF at the position of the tunnel barrier
as follows:
\begin{eqnarray}
\mathcal{G}_{j}(\tau) =  -\sum_{\alpha}\langle T_{\tau}\psi_{\alpha,j}^{(0)}(\tau)\hat{F}_{j}(\tau)\hat{F}_{j}^{\dagger}(0)\psi_{\alpha,j}^{(0)\dagger}(0)\rangle~.
\end{eqnarray}
Since the operators $\psi_{\alpha,j}^{(0)}$ and $\hat{F}_{j}$ are
decoupled, the GFs are factorized as $\mathcal{G}_{j}(\tau)=G_{0,j}(\tau)K_{j}(\tau)$,
where $G_{0,j}(\tau)=-\nu_{0,j}\pi T/\sin(\pi T\tau)$ being a free
fermion's GF, $\nu_{0,j}$ is the DoS in the dot computed
in the absence of renormalization effects associated with electron-electron interaction
\cite{com1}.
Therefore, all effects of interaction and scattering are accounted for by the correlator  $K_{j}(\tau)=\langle T_{\tau}\hat{F}_{j}(\tau)\hat{F}_{j}^{\dagger}(0)\rangle$
\cite{MAtheory}.

The current is calculated in the linear response regime. We neglect the resistance of  the metallic QDs and assume that the voltage difference $\Delta {\cal V}_{\rm th}$ arises across the tunnel barrier between two QDs. The transport coefficients, namely, the electric conductance  $G$ and the thermoelectric coefficient $G_{T}$ (measured independently) define the Seebeck effect quantified in terms of the thermoelectric power (TP) $S$:
\[
G_{T}=\frac{\partial I_{\rm sd}}{\partial \Delta T},\;\;\; G=\frac{\partial I_{\rm sd}}{\partial {\cal V}},\;\; S=-\left.\frac{\Delta {\cal V}_{\rm th}}{\Delta T}\right|_{I_{\rm sd}=0}=\frac{G_{T}}{G}.
\]
Plugging the DoS $\nu_1(\epsilon)$ and $\nu_2(\epsilon)$ in (\ref{tuncur}), we express the electric conductance as follows:
\begin{eqnarray}
G=\!\frac{\pi}{2} G_{C} T\!\!\int_{-\infty}^{\infty}\!\!\!\!\!\!\! d\tau\frac{K_{1}\left(\frac{1}{2T}+i\tau\right) K_{2}\left(\frac{1}{2T}-i\tau\right)}{\cosh^{2}(\pi T\tau)}.
\label{thercond30}
\end{eqnarray}
The thermoelectric coefficient $G_T$  is given by
\begin{eqnarray}
\!\!\!\! G_{T}=\frac{iG_{C}}{8e}\!\!\!\int_{-\infty}^{\infty}\!\!\!\!\!\!d\tau\frac{W\left[K_{1}\left(\frac{1}{2T}+i\tau\right),\; K_{2}\left(\frac{1}{2T}-i\tau\right)\right]}{\cosh^{2}(\pi T\tau)}~.
\label{thercond31}
\end{eqnarray}
Both quantities are expressed in terms of the correlators $K_j$ analytically continued to real time. Here we introduce a short-hand notation
\begin{equation}
G_{C}=2\pi e^{2}\nu_{01}\nu_{02}|t|^{2}
\end{equation}
for the conductance of the tunnel (central) area between two QD-QPC devices.
The Wronski determinant $W$ is defined as follows: 
\begin{eqnarray}
&&W\left[K_{1}\left(\frac{1}{2T}+i\tau\right),\; K_{2}\left(\frac{1}{2T}-i\tau\right)\right]\nonumber\\
&=&\left|{\displaystyle \begin{array}{cc}
{\displaystyle K_{1}\left(\frac{1}{2T}+i\tau\right)} & {\displaystyle K_{2}\left(\frac{1}{2T}-i\tau\right)}\\
{\displaystyle \frac{d}{d\tau}K_{1}\left(\frac{1}{2T}+i\tau\right)} & {\displaystyle \frac{d}{d\tau}K_{2}\left(\frac{1}{2T}-i\tau\right)}
\end{array}}\right|~.\label{wronskian}
\end{eqnarray}
Notice that the Wronski determinant (wronskian) is zero when two functions are linearly
dependent. By integrating by parts the integral containing the wronskian, we obtain
equation for the thermo-electric coefficient $G_T$ which we use in our calculations (see Section \ref{Sec4}):
\begin{eqnarray}
G_{T}\!&=&\!\frac{iG_{C}}{8e}\!\!\!\int_{-\infty}^{\infty}\!\!\!\!\!\!\!\! d\tau
\!\!\left[\! -2\pi T\!\frac{\sinh(\pi T\tau\!)}{\cosh^{3}(\pi T\tau\!)}\!K_{1}\!\!\left(\!\frac{1}{2T}\!+i\tau\!\!\right)\! K_{2}\!\!\left(\!\frac{1}{2T}\!-i\tau\!\!\right)\right.\nonumber \\
&&\!\!\! +\left.\!\frac{2}{\cosh^{2}(\pi T\tau)}\! K_{1}\!\!\left(\!\frac{1}{2T}\!+i\tau\!\right)\!\!\left\{ \!\frac{d}{d\tau}\! K_{2}\!\!\left(\!\frac{1}{2T}-i\tau\!\right)\!\right\}\!\right].\label{cond3d}
\end{eqnarray}
Comparison of transport coefficient calculated by Eq. (\ref{thercond30}-\ref{thercond31})
with Matveev-Andreev limiting cases in which either $K_{2}$ or $K_{1}$ are $\tau$ - independent \cite{MAtheory} can be straightforwardly done with a help of Eq.(\ref{wronskian}).

\section{Main Results}\label{Sec4}

Using the general formalism sketched in the previous Section we proceed straightforwardly 
to calculation of the thermo-electric coefficients and 
discussion of our main results. The setup (Fig. \ref{f.1-1}) allows to engineer various (Fermi- and Non-Fermi-Liquid) states
connected through the weak link. We label and discuss three important limiting cases one by one in the Subsections below. We assume in all calculations that the condition $T\ll {\rm min}[E_{C,1},E_{C,2}]$
is fulfilled \cite{MAtheory}.

\subsection{Fermi Liquid vs Fermi Liquid }

The first (trivial) case corresponds to turning off one of the QPCs on both left and right 
QD-QPC devices. We assume for illustration purposes that the QPC B and QPC C
are turned off. As a result, the electric circuit consists of QPC A - QD1 -- QD2 - QPC D.
Both left and right devices are tuned to FL regime separated by the weak link. Each of the FL states originates from a single channel Kondo strong coupling fixed point \cite{MAtheory}. 

The correlators $K_j\left(\tau\right)$ are obtained by using a perturbation theory
in $|r_j|$ \cite{MAtheory} and are given by
\begin{eqnarray}
K_{j}\!\left(\tau\right)&=&\!\left(\frac{\pi^{2}T}{\gamma E_{C,j}}\right)^{2}\!\!\!\frac{1}{\sin^{2}\left(\pi T\tau\right)}\left[1-2\gamma\xi|r_{j}|\cos\left(2\pi N_{j}\right)\right.\nonumber\\
&&\left. +4\pi^{2}\xi\gamma|r_{j}|\frac{T}{E_{C,j}}\sin\left(2\pi N_{j}\right)\cot\left(\pi T\tau\right)\right]~.
\label{KFL}
\end{eqnarray}
{\color{black} where $\xi=1.59$ is a numerical constant \cite{MAtheory},
$|r_1|=|r_A|$ and $|r_2|=|r_D|$.}
Substituting  $K_j(\tau)$ into Eqs. (\ref{thercond30}) and (\ref{thercond31}) we get the
linear response equations for the differential conductance and the thermoelectric coefficient:
\begin{eqnarray}
G&=&\frac{8\pi^{8}G_{C}}{15\gamma^{4}}\frac{T^{4}}{E_{C,1}^{2}E_{C,2}^{2}}~,\\
\label{GFLFL}
G_{T} & =&-\frac{32\pi^{10}\xi G_{C}}{35 e \gamma^{3}}\frac{T^{5}}{E_{C,1}^{2}E_{C,2}^{2}}\!\!\sum_{j=1,2}\!\!|r_{j}|\frac{\sin\left(2\pi N_{j}\right)}{E_{C,j}}~.
\label{GTFLFL}
\end{eqnarray}
The resulting thermopower (Seebeck coefficient) is obeying the FL equation:
\begin{equation}
S=-\frac{12}{7}\frac{\pi^{2}\gamma\xi}{e}\sum_{j=1,2}|r_{j}|\frac{T}{E_{C,j}}\sin\left(2\pi N_{j}\right)~.
\label{S_FLFL}
\end{equation}
The thermoelectric transport regime of two weakly coupled Fermi liquids can be used
for benchmarking and calibration necessary for quantifying the deviations 
from the FL transport properties.

\subsection{Fermi Liquid vs Non-Fermi Liquid}
The second important case describes the FL one-channel Kondo state in the left device
(QPC A is switched on and QPC B is switched off) weakly connected to the Non-Fermi-liquid state
corresponding to the two-channel Kondo effect when both QPC C and QPC D of the right device are turned on. In addition, we assume that the reflection amplitudes in both QPCs (C and D) are fine-tuned to the symmetric regime
$|r_C|=|r_D|$ necessary for protection of the NFL state.
In reality, small detuning of the reflection amplitudes suppresses the NFL state at sufficiently low temperature $T^R_{\rm min}$ determined by the asymmetry of $r$'s ($T^R_{\rm min}\propto ||r_C|-|r_D||^{2}E_{C2}$ with $||r_C|-|r_D||\rightarrow 0$) \cite{nkk_2010}. To avoid a collapse to FL ground state we assume that the measurement's temperature $T$ is higher than $T^R_{\rm min}$.

We use the same FL equation for $K_1$ (which depends linearly on 
$|r_1|=|r_A|$) as in the previous Subsection. The perturbative expansion for $K_2$ in terms of 
small $|r_2|=|r_C|=|r_D|$, however,
starts with the second-order term. The reason is that the fluctuations of the iso-spin mode in the right device are not suppressed  at low energies by the charging energy.
In contrast to it, the iso-spin projection is fixed in the left device and its  fluctuations are completely frozen. We point out that the left-right symmetry is explicitly violated in this setup by the construction. Following Matveev-Andreev calculations (see \cite{MAtheory} for details) we use the following equation for $K_2$:
\begin{eqnarray}
&& K_{2}\left(\tau\right)=\frac{\pi^{2}T}{2\gamma E_{C,2}}\frac{1}{|\sin\left(\pi T\tau\right)|}\nonumber\\
&&\!\!\times\left[1-\frac{8\gamma}{\pi^{2}}|r_{2}|^{2}\sin\left(2\pi N_{2}\right)\ln\!\left(\frac{E_{C,2}}{T}\right)\!\ln\tan\!\left(\!\frac{\pi T\tau}{2}\!\right)\!\right]~.
\label{KNFL}
\end{eqnarray}
Plugging Eq. (\ref{KNFL}) in Eqs. (\ref{thercond30}) and (\ref{thercond31}) we get the transport coefficients:
\begin{eqnarray}
G &=&\frac{3\pi^{7}G_{C}}{32\gamma^{3}}\frac{T^{3}}{E_{C,1}^{2}E_{C,2}}~,\\
\label{GFLNFL}
G_{T} &=&-\frac{\pi^{4} G_C}{e\gamma^{2}}\!\frac{T^{3}}{E_{C,1}^{2}E_{C,2}}\left[\frac{\pi^5\xi}{16}\frac{T}{E_{C,1}}|r_{1}|\sin\left(2\pi N_{1}\right)\right.\nonumber\\
&&\left.+\frac{16}{25}\ln\!\left(\frac{E_{C,2}}{T}\!\right)\!|r_{2}|^{2}\sin\left(2\pi N_{2}\right)\right]~.
\label{GTFLNFL}
\end{eqnarray}
The thermopower in the lowest order of $|r_{j}|$ contains two competing terms
\begin{eqnarray}
S&=&-\frac{32\gamma}{3e\pi^{3}}\left[\frac{\pi^5\xi}{16}\frac{T}{E_{C,1}}|r_{1}|\sin\left(2\pi N_{1}\right)\right.\nonumber\\
&&\left.+\frac{16}{25}\ln\!\left(\frac{E_{C,2}}{T}\!\right)\!|r_{2}|^{2}\sin\left(2\pi N_{2}\right)\right]~.
\label{S_FLNFL}
\end{eqnarray}
The crossover line separating dominant FL contribution to the TP from the dominant NFL contribution is defined as follows
\begin{equation}
\frac{256}{25\xi\pi^{5}}\ln\left(\frac{E_{C,2}}{T^\ast}\right)\frac{E_{C,1}}{T^\ast}=\frac{|r_{1}|}{|r_{2}|^2}~.
\label{specialT0}
\end{equation}
If $T\ll T^\ast$ pronounced NFL behaviour of the Seebeck effect is predicted.
In the opposite limit,  $T\gg T^\ast$, the FL regime with the weak NFL corrections
is expected.

{\color{black} As it was discussed in \cite{Furusaki_Matveev}, the Non-Fermi liquid behaviour is manifested in $T$-dependence of both electric conductance $G$ and thermoelectric coefficient $G_T$. The $\propto T^3$ behaviour of the conductance is originating from the $\propto T^2$ FL behaviour \cite{Furusaki_Matveev} and $\propto T$ NFL behaviour arises from the Anderson orthogonality catastrophe \cite{catastrophe,catastrophe1,Mahan}. The orthogonality 
related to change of the boundary condition in QD results in an appearance of 
a resonance scattering phase $\delta =\pi/2$ \cite{Furusaki_Matveev} and leads to a power-law suppression of the density of states $\nu_2(\epsilon)\propto \epsilon$.
The same density of states effects in thermoelectric coefficient justify the log-$T$ behaviour of the thermopower.}

\subsection{Non-Fermi Liquid vs Non-Fermi Liquid}
The third important case describes two Non-Fermi Liquid  
states connected by the weak link.
This case can be engineered by turning on all four QPCs in the left and right devices with a simultaneous fine-tuning of the reflection amplitudes
as $|r_1|=|r_A|=|r_B|$ and $|r_2|=|r_C|=|r_D|$.
As it has been explained above, the small detuning of each pair of the reflection amplitudes in the left and right devices results in an emergence of two characteristic 
temperature scales $T^L_{\rm min}$ and $T^R_{\rm min}$ \cite{nkk_2010}. We discuss therefore the thermoelectric transport at $T\gg {\rm max}[T^L_{\rm min},T^R_{\rm min}]$.

Substituting the Non-Fermi Liquid correlators for $K_{j}\left(\tau\right)$ [Eq. (\ref{KNFL})] into general equations for the transport coefficients, we get:
\begin{eqnarray}
&&\!\!\! G=\frac{\pi^{4}G_{C}}{6\gamma^{2}}\frac{T^{2}}{E_{C,1}E_{C,2}}~,\\
\label{GNFLNFL}
&&\!\!\!\!\! G_{T}\!\! =\! -\frac{3\pi^{3}G_{C}}{32 e\gamma}\!\frac{T^{2}}{E_{C,1}E_{C,2}}\!\!
\sum_{j=1,2}\!\!\!|r_{j}|^{2}\!\ln\!\left(\!\!\frac{E_{C,j}}{T}\!\!\right)\!\sin\!\left(2\pi N_{j}\!\right).
\label{GTNFLNFL}
\end{eqnarray}
We emphasize that the iso-spin fluctuations in both devices, being not suppressed by the charging energy, play crucial role and determine the NFL behaviour of the thermopower:
\begin{equation}
S=-\frac{9\gamma}{16e\pi}\!\sum_{j=1,2}|r_{j}|^{2}\ln\left(\frac{E_{C,j}}{T}\right)\sin\left(2\pi N_{j}\right)~.
\label{S_NFLNFL}
\end{equation}
{\color{black} Actually, pronounced Non-Fermi Liquid behaviour of thermopower is associated
with  two orthogonality catastrophes \cite{catastrophe,catastrophe1,Mahan} emerging simultaneously  in the left and the right quantum devices 
$\nu_1(\epsilon)\sim\nu_2(\epsilon)\propto\epsilon$. As a result, both densities of states are suppressed due to change of the boundary conditions and emergence of two resonance scattering phases $\delta =\pi/2$.
}
\begin{figure}[t]
\includegraphics[width=50mm]{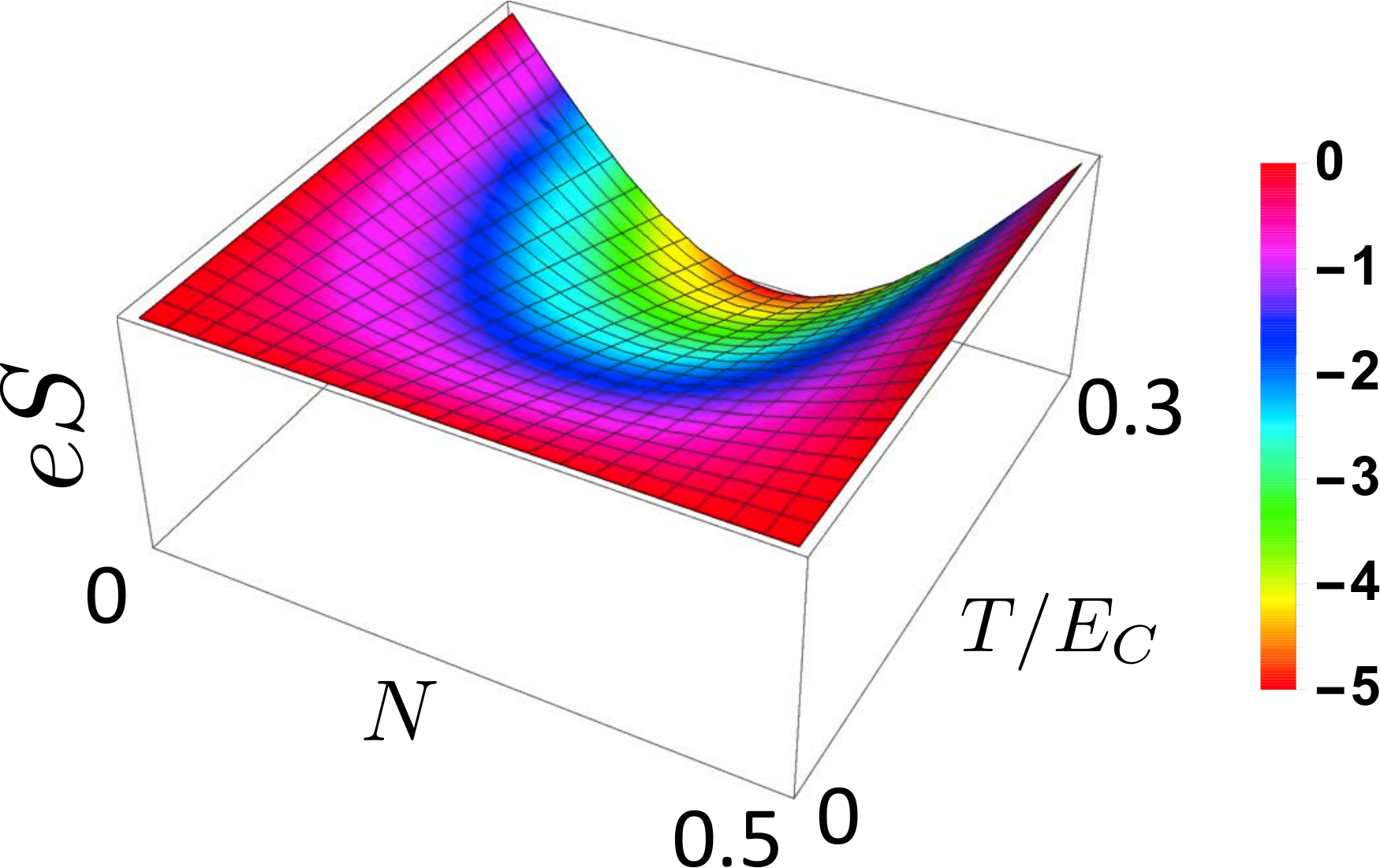}\hspace*{5mm}
\includegraphics[width=30mm]{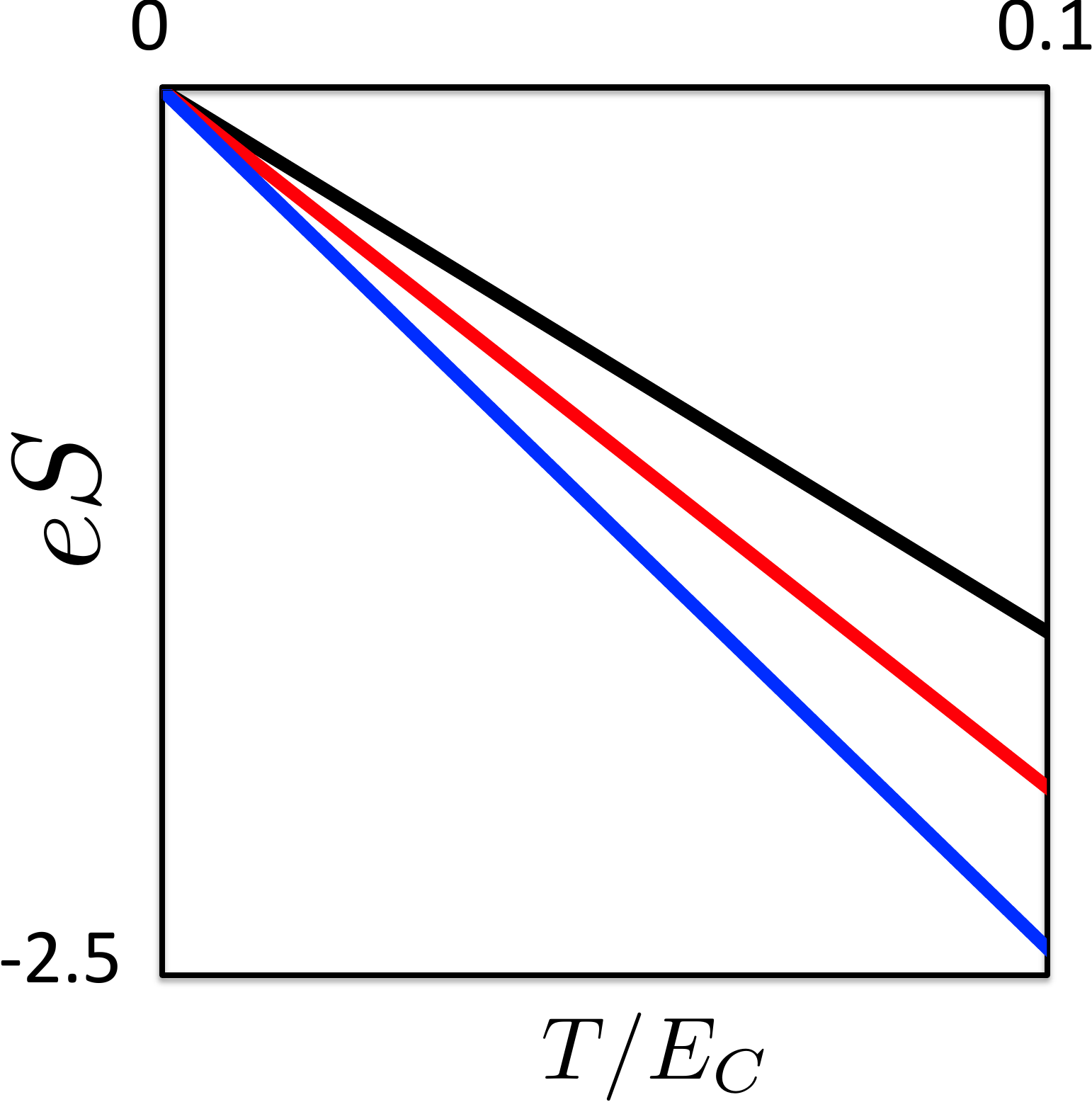}\\
\includegraphics[width=50mm]{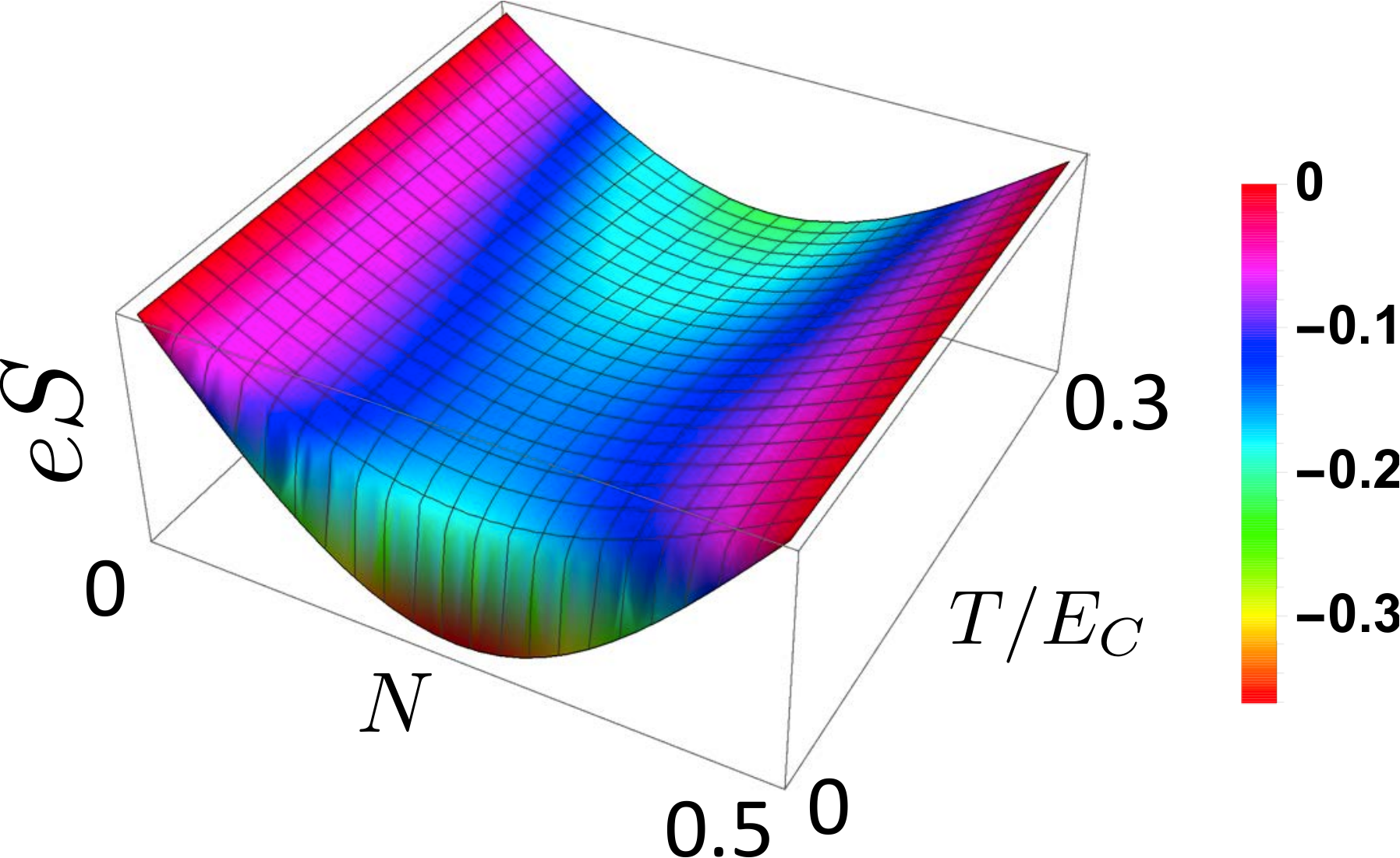}\hspace*{5mm}
\includegraphics[width=30mm]{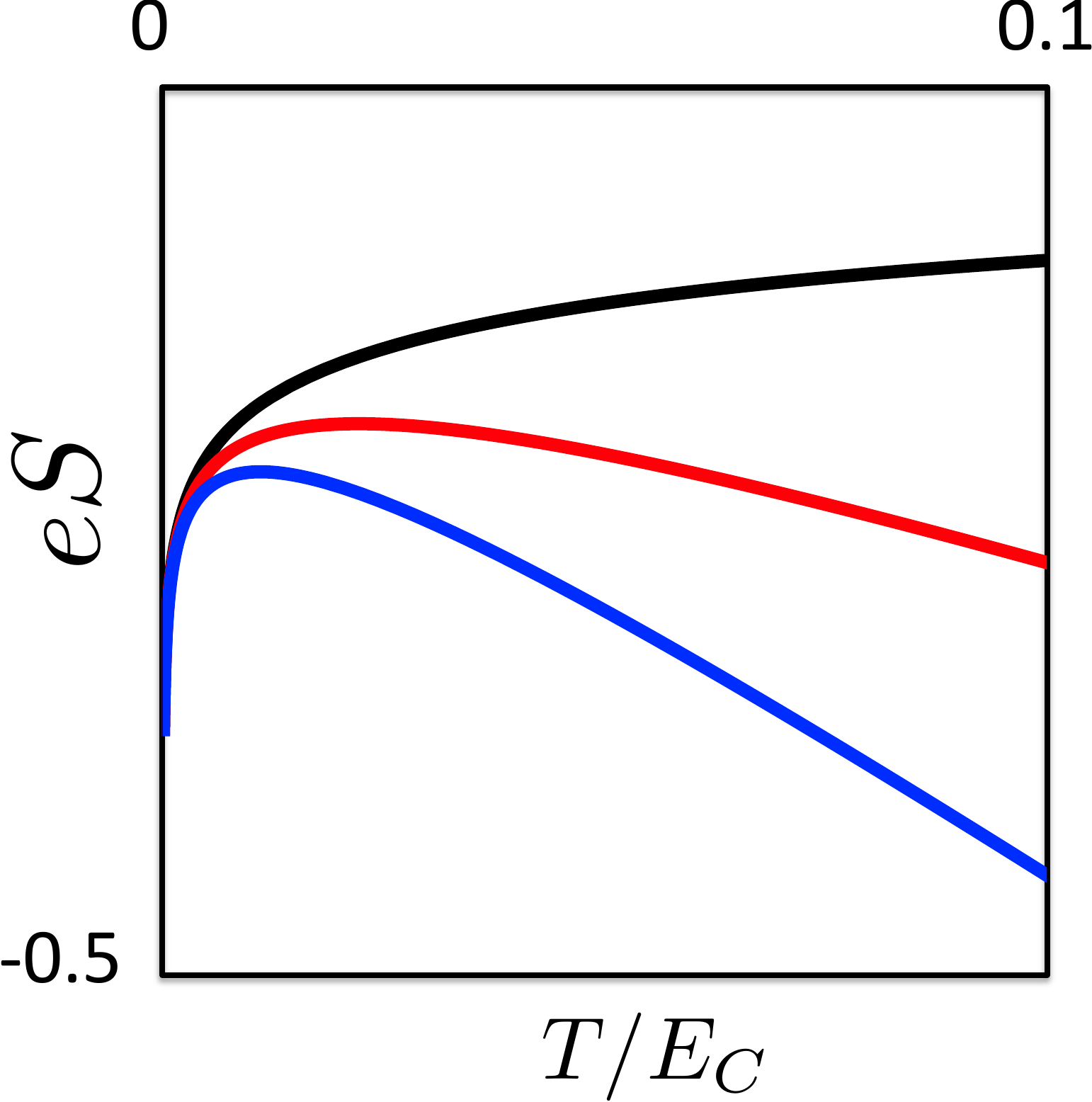}\\
\includegraphics[width=50mm]{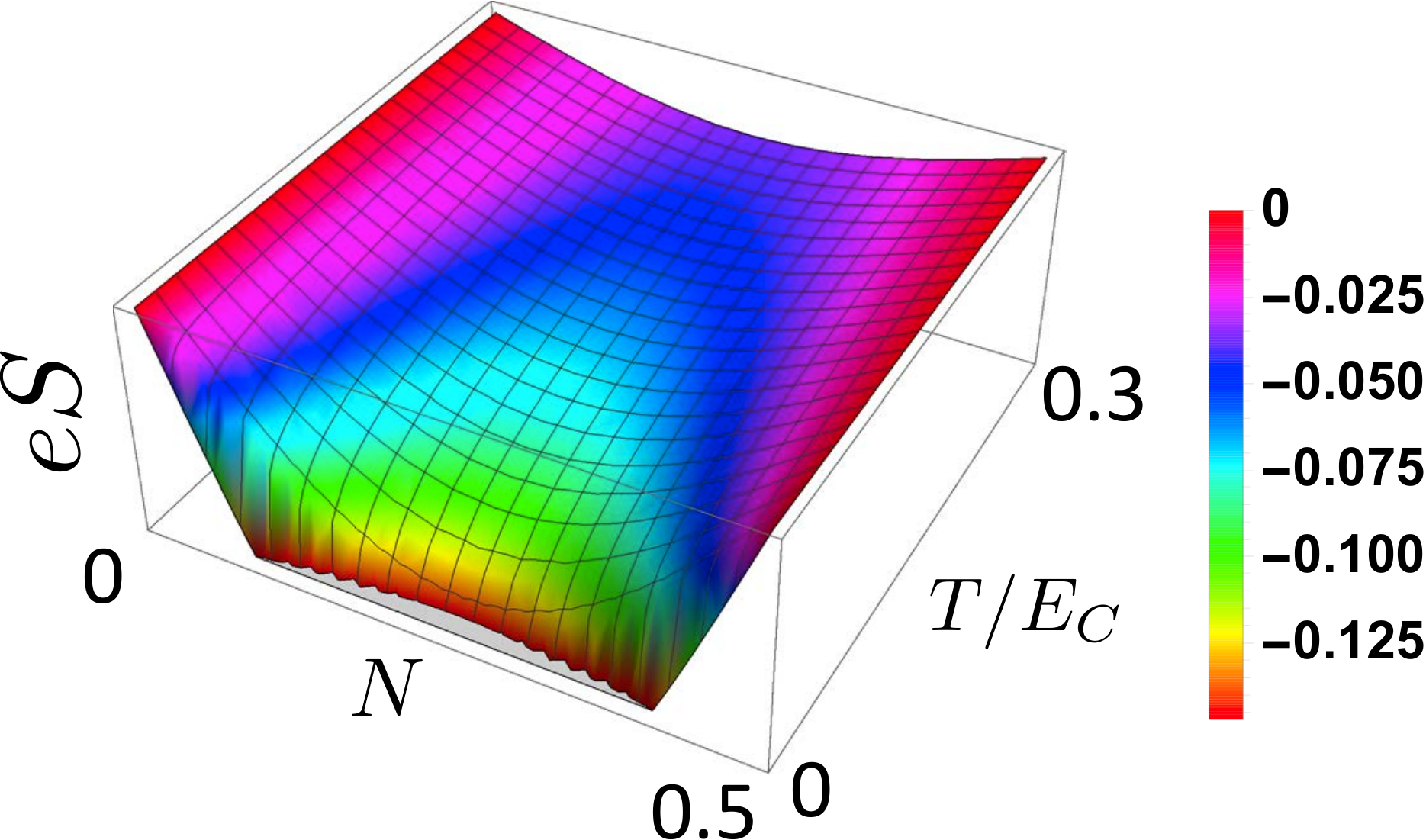}\hspace*{5mm}
\includegraphics[width=30mm]{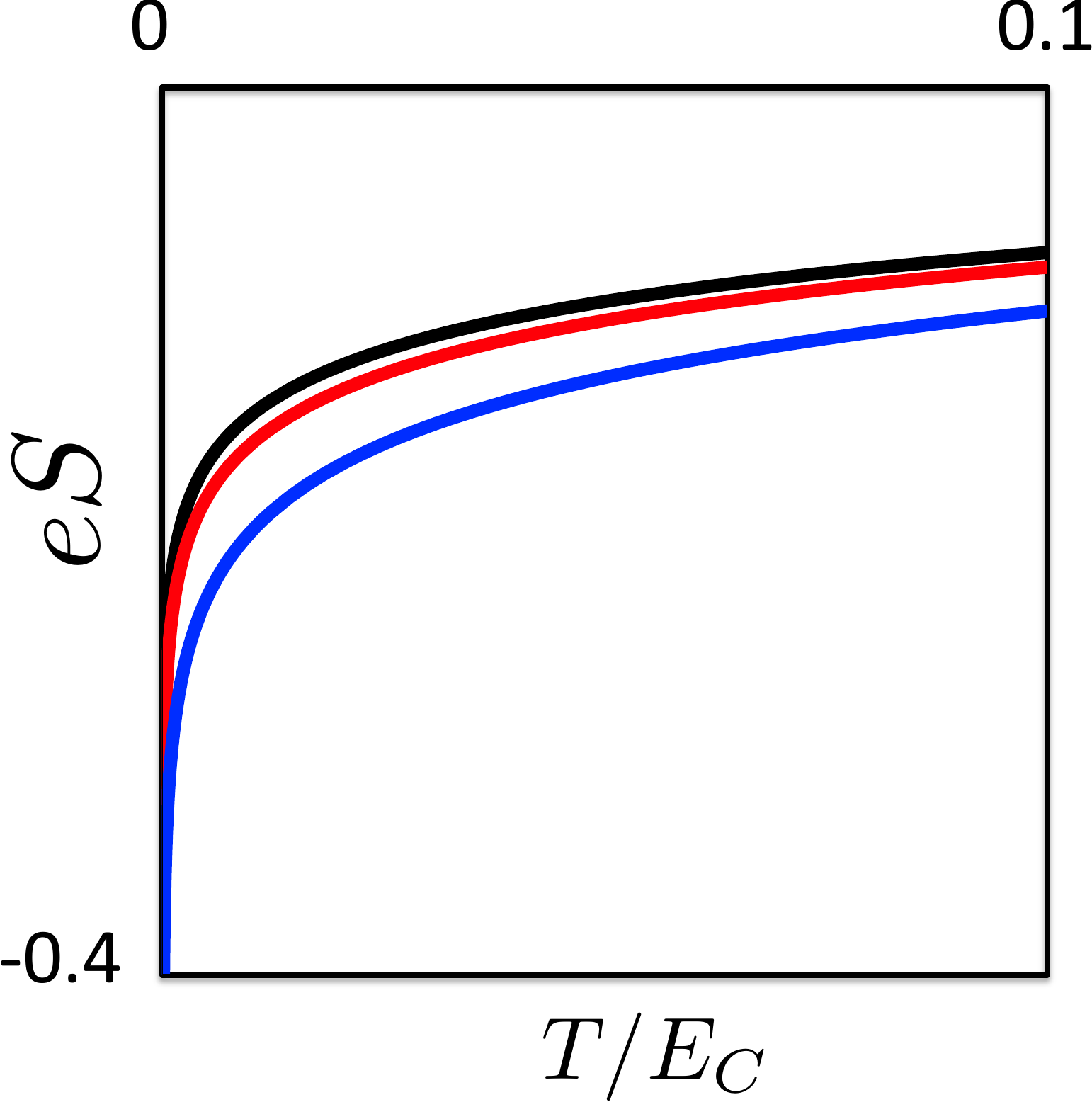}
\caption{(Color online) Left panels: Plots of thermopower $eS$
as a function of $N=N_{1}=N_{2}$ and temperature $T/E_C$ ($E_{C1}=E_{C2}=E_{C}$) 
with $|r_{1}|=\sqrt{0.001}, \: |r_{2}|=\sqrt{0.1}$ a) for the FL-FL regime (upper panel). b) for the FL-NFL regime (central panel). c) for the NFL-NFL regime (lower panel). 
Right panels: Plots of the minimum of thermopower $eS$ as a function
of temperature $T/E_C$ for different left/right asymmetries $|r_1/r_2|=0.01$ (black curve), 
$|r_1/r_2|=0.3$ (red curve), and $|r_1/r_2|=0.6$ (blue curve) with $|r_{2}|=\sqrt{0.1}$ 
for the regimes of the left panels correspondingly.}
\label{fSNT} 
\end{figure}

\section{Discussion}\label{Sec5}
The summary of the qualitative and quantitative behaviours of the TP in three
important Fermi Liquid and Non-Fermi Liquid regimes is illustrated in the Fig. \ref{fSNT}.

\subsection{Fermi Liquid to Non-Fermi Liquid crossover}

The 3D plots (left panels of Fig. \ref{fSNT}) show the thermopower dependence on the gate voltage $N=N_1=N_2$ and the temperature $T$. There is a significant difference between three 
different regimes. i) The upper plot (left panel) shows of a minimum of TP at 
$N=1/4$ with a depth increasing with increase the temperature. The behaviour
$S_{\rm min}\propto T$ is characteristic for the FL regime (see the right panel of the upper plot where the temperature dependence of $S_{\rm min}$ is shown; the black/red/blue curves describe
the evolution of $S$ with decreasing of transparency of the left device (see Fig.\ref{fSNT} and figure caption for detail).
ii) The central part of Fig. \ref{fSNT}
describes the Seebeck effect on a weak link between Fermi- and Non-Fermi liquids. There is visible competition between two minima: one minimum is associated with the FL properties
(linear in $T$ dependence of $S_{\rm min}$), while the second minimum is characterized by
the $\text{log-}T$ scaling arising at low temperatures. The FL-NFL competition manifests itself in a non-monotonous temperature dependence of $S_{\rm min}$ [cf. the right panel of the central part of the Fig. \ref{fSNT} where one sees the $S_{\rm min}\propto\ln(T^\ast /T)$ at  $T\ll T^\ast $ and  $S_{\rm min} \propto T/T^\ast$ at $T\gg T^\ast$ in accordance with Eq. (\ref{specialT0})]. iii) The lower panel of Fig. \ref{fSNT} shows the TP behaviour at 
the weak link between two NFL regimes. Since in both NFL states of two-channel Kondo origin there exist strong fluctuations of the iso-spin, the characteristic behaviour of TP
shows pronounced NFL $\text{log-}T$ scaling. The $S_{\rm min}$ temperature dependence is monotonic $S_{\rm min}\propto \ln T$ providing an evidence of the NFL thermo-transport behaviour
\cite{com2}.


Summing up, we have demonstrated that engineering two weakly coupled nano-devices
allows full control on the crossover between Fermi- and Non-Fermi Liquid thermoelectric transport with three spectacular regimes: Fermi-liquid, competing  Fermi- and 
Non-Fermi Liquids and pronounced Non-Fermi Liquid.

\subsection{Prospectives: from weak link through strong coupling to multi-channel and multi-impurity Kondo physics}

\begin{figure}[t]
\includegraphics[width=7cm]{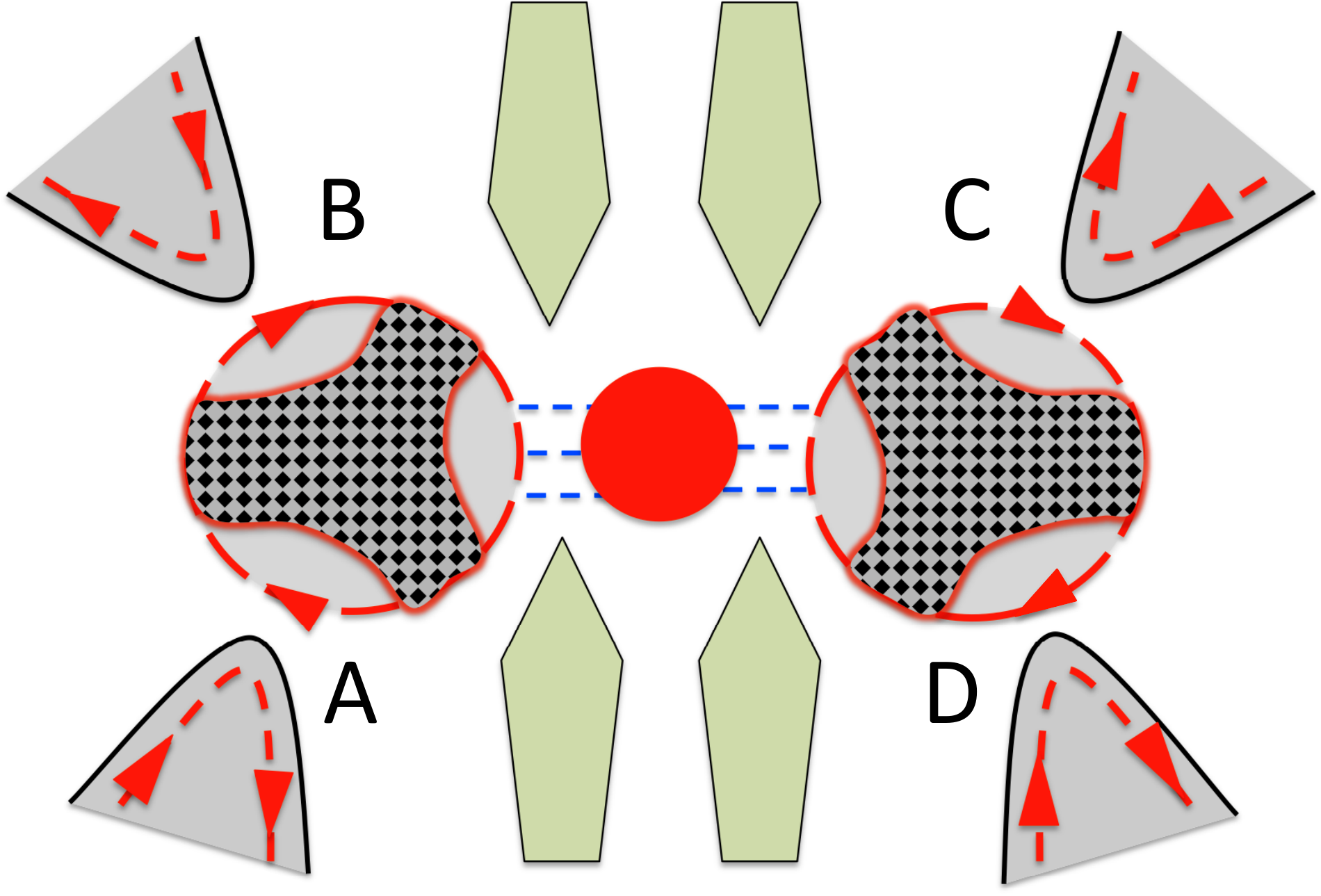} \caption{(Color online) Sketch of "Bell test" thermoelectric experiment: central red circle denotes "hot" QD weakly coupled by left and right tunnel barriers (blue dashed lines) to the "cold" QD-QPC devices. All other notations are identical to those used in Fig. \ref{f.1-1}.}
\label{f.1-4} 
\end{figure}

The simplicity of the model discussed in our paper is based on a trivial assumption:
both the temperature and voltage drops are occurring across the weak link (tunnel barrier).
As a result, we can characterize each of two quantum systems (to the left and to the right
from the tunnel barrier) by the temperature and chemical potential assuming that 
each system is at equilibrium. Besides, the zero-current steady state 
condition implemented for the Seebeck effect measurements gives 
a direct access to the off-diagonal transport
coefficient $G_T$ by avoiding a necessity to compute the heat current \cite{Luttinger}.

{\color{black} A straightforward generalization of the model considered above could be achieved by adding to a setup a central island connected to the left and right devices by two tunnel barriers (see Fig. \ref{f.1-4}). The central island (metallic quantum dot with a continuous spectrum) is artificially heated to guarantee a temperature drop across two
tunnel barriers. The hot electrons therefore can be emitted to the left and right 
devices the same way as photons are emitted in famous Bell test experiments 
\cite{Bell,Belltest}. It opens a possibility to measure the current-current correlation functions in two- and four- terminal geometries. In addition, the device shown in Fig. \ref{f.1-4}
is suitable for investigation of a heat quantization 
\cite{heatquantization} and heat Coulomb blockade \cite{heatblockade}, universality  of thermo-conductance fluctuations
\cite{heatfluctuations} and heat entanglement.}

\begin{figure}[t]
\includegraphics[width=75mm]{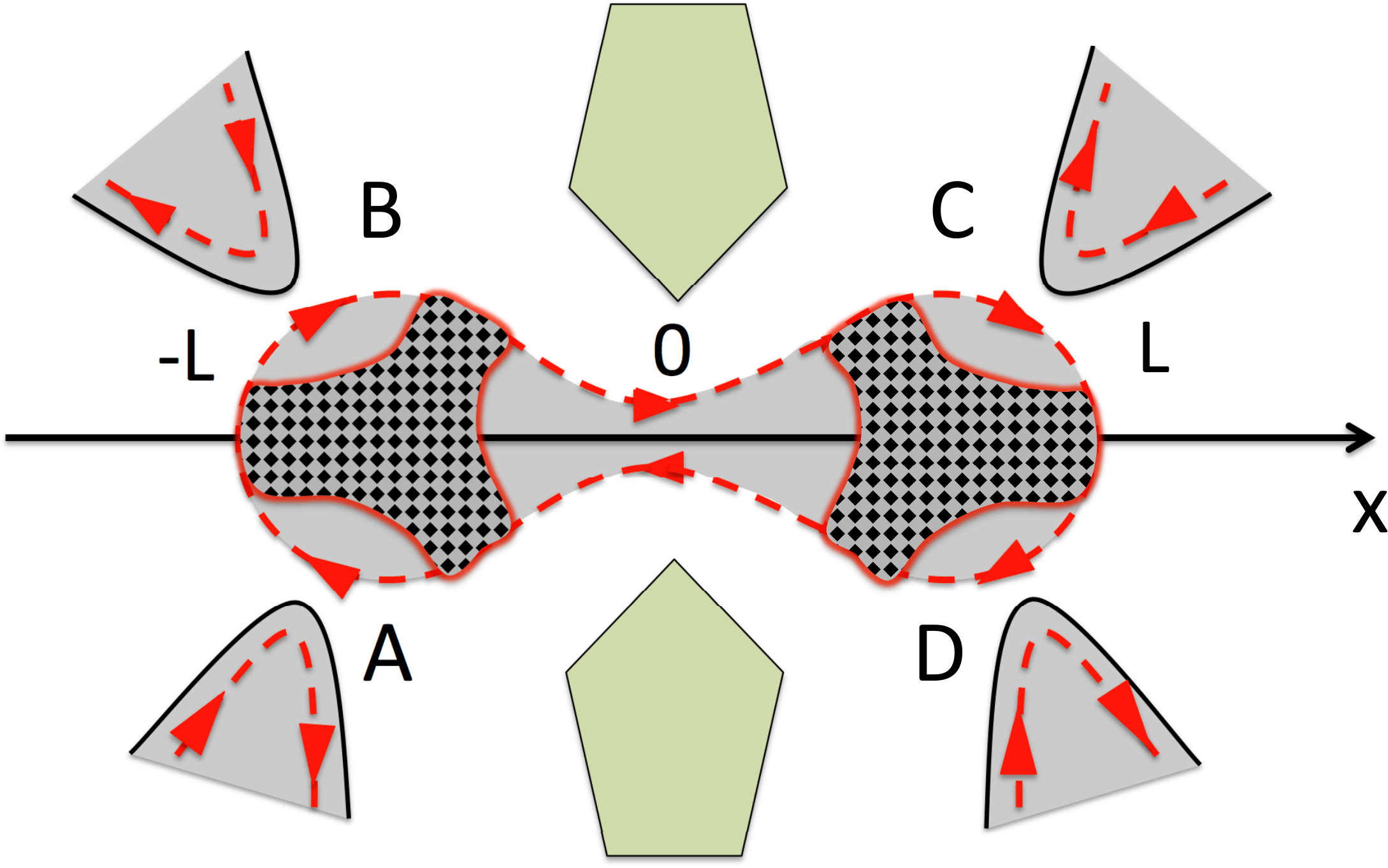} \caption{(Color online) Schematic representation of a hybrid metal-semiconductor device engineering a strong coupling of two quantum dots. A peanut-shape 2DEG area is formed by opening a conducting channel between dots
(central quantum point contact at $x=0$) by the split gate (light green boxes).
All other notations are the same as on Fig. \ref{f.1-1}. The hybrid device can be
fine tuned to different multi-channel ($1 \to 4$ channel) Kondo regimes
of a single quantum impurity (fully transparent central QPC) and different regimes
of multi-channel ($1 \to 2$ channel) double quantum impurity setups (see detailed discussion in the Section \ref{Sec5}). Fine tuning the split gates at positions A-B (at $x=-L$) and/or C-D
(at $x=L$) to weak (tunnel) regime and heating one of the peripheral contacts  allows to
measure a thermo-electric transport on a weak link between one-to-three channel Kondo regimes.}
\label{f.1-3} 
\end{figure}

Another interesting model arises if we replace a weak link between two
quantum devices by a strong link engineered for example by an additional QPC \cite{tbp}.
If the central QPC (see Fig. \ref{f.1-3}) is widely opened and therefore fully transparent, 
there exists unique edge state which merges a double dot device into a single peanut-shaped QD.  Therefore, turning on the QPCs  one-by-on the  by the split gates provides a possibility to investigate the charge transport through 1-4 channel Kondo devices.
Besides, if one of four QPCs (A-D) is converted to the tunnel junction with the temperature 
and voltage drop across it, the thermoelectric transport to the NFL states
associated with two-channel \cite{2CK_experiment_nature2015} and three-channel \cite{3CK_experiment} Kondo effects become accessible.
While the NFL physics emerging in the vicinity of the strong coupling fixed
point of the two-channel Kondo is described by Majorana ($Z_2$) fermions, the NFL
physics of the three-channel Kondo  is governed by $Z_3$ parafermions \cite{3CK_experiment}. The Seebeck coefficient (thermoelectric power) provides information on change of the Coulomb energy
across the dot associated with voltage drop per fixed temperature drop and therefore
gives some clue about the entropy flux through the nano-device \cite{Beni}. 
Thus, the thermoelectric transport measurements of the N-channel Kondo devices
might shed some light on the $Z_N$ parafermion excitations \cite{CFT}.

{\color{black} Change of the central QPC transparency can be 
controlled by the split gate (see Fig. \ref{f.1-3}) . By squeezing the constriction, the QPC can be fine-tuned to a single-mode low reflection regime}. The peanut-shaped central QD 
\cite{LeHur2003,peanut} is split into two parts (two QDs with quantized charge) and therefore operates in the regime described by the two-impurity Kondo model. By construction, two impurities are attached to several orbital channels modelled by QPCs A-D. 
{\color{black} The unstable strong coupling fixed point of } the  
two-impurity single channel Kondo model can be mapped to the two-channel Kondo model \cite{Jones_Varma,Gan_95,Logan2012} and therefore results in the NFL properties.
Similarly to single-impurity multi-channel device, by converting one of A-D QPCs to the tunnel junction regime, one can investigate the thermoelectric transport
through the multi-impurity multi-channel Kondo devices. High tunability of the QPCs will provide an access to various (Fermi and Non-Fermi liquid) fixed points of the quantum systems.

\section{Summary and Conclusions}\label{Sec6}

We propose a theoretical model describing a hybrid quantum device consisting of two 
quantum system coupled through a weak tunnel link. Each of the quantum systems in turn
consists of a quantum dot strongly coupled to several quantum point contacts. The idea of
the setup is based on existing experimental devices designed for investigation of
multi-channel Kondo physics \cite{current_heat, 2CK_experiment_nature2015, 3CK_experiment}. Being inspired by the new measurements of the quantum charge states \cite{2CK_experiment_nature2015}, 
we suggest to utilize similar geometry of Integer Quantum Hall nano-structures for measurements of thermo-electric transport. The "weak link" quantum system suggested in our paper demonstrates
(depending on fine-tuning the external parameters, such as gate voltages applied to the split gates controlling the QPCs) three significantly different regimes, such as, weak link between:
i) two Fermi Liquid states, ii) Fermi and Non-Fermi-liquid states and iii) two
Non-Fermi Liquid states. In order to demonstrate the signature of the Non-Fermi Liquid physics in thermo-electric transport properties, we calculated the Seebeck coefficient
(TP) perturbatively in QPC reflection amplitudes for all three generic cases. It was shown that the FL-FL setup can be used for calibration of the Fermi-liquid
behaviour of the TP. The system of weakly coupled FL-NFL demonstrates pronounced
competition between two types of contribution to the Seebeck coefficient: i) linear in $T$ characteristic for the FL and ii) $\text{log-}T$ behaviour specific for the NFL.
One can use this setup for determining the crossover temperature which separates two
different regimes. The benchmark of the crossover effect is the non-monotonicity of 
the TP minimum located at $N_1=N_2=1/4$ as a function of temperature.
The most spectacular illustration of the NFL behaviour of the Seebeck coefficient is
demonstrated on the weak link between two NFL states. It shows the pronounced
$\text{log-}T$ dependence of the TP at sufficiently wide temperature interval.
The central idea of engineering of the NFL states is based on high tunability
of the multi-channel Kondo physics in QD-QPC nano devices \cite{2CK_experiment_nature2015, 3CK_experiment}. {\color{black} The challenge of  proposal for new experiments is to boost efforts for understanding the  thermoelectric transport in Non-Fermi Liquid regimes emerging due to strong electron correlations and resonance  scattering. The setup then can be used as a controllable playground for studying the thermo-electric phenomena in nano-devices.} 

\section*{Acknowledgements} 

We are grateful to S. Jezouin, L.W. Molenkamp, F. Pierre for illuminating discussions,
A. Komnik for cooperation on the initial stage of this project and critical reading of the manuscript and A. Parafilo for useful comments on the structure of the paper. We acknowledge B.L. Altshuler's idea \cite{privcom} suggesting to use thermoelectrics for Bell's test measurements. T.K.T.N. acknowledges support through the Associate Program of ICTP.  This research in Hanoi is funded by Vietnam National Foundation for Science and Technology Development (NAFOSTED) under grant number 103.01-2016.34. 



\begin{thebibliography}{10}
\bibitem{com1} Such experimental measurement can be done by complete opening the QPC.
\end{thebibliography}
\end{document}